\makeatletter
\declare@file@substitution{revtex4-1.cls}{revtex4-2.cls}
\makeatother

\documentclass[]{aastex62}

\usepackage{amsmath}
\usepackage{mathrsfs}
\usepackage[caption=false]{subfig}
\usepackage{threeparttable}
\usepackage{booktabs}
\usepackage{longtable}

\usepackage{CJKutf8}

\shorttitle{Event rate in KN/GRBs}
\shortauthors{Li et al.}
\begin{document}

\title{The Cosmic Star Formation History: Insights from Kilonova-Associated Gamma-Ray Bursts}

\correspondingauthor{Qin-Mei Li and Sheng-Bang Qian}
\email{qinmli@163.com, qiansb@ynu.edu.cn}

\author{Qin-Mei Li}
\affiliation{Department of Astronomy, School of Physics and Astronomy, Yunnan University, Kunming 650091, China}

\author{Qi-Bin Sun}
\affiliation{Department of Astronomy, School of Physics and Astronomy, Yunnan University, Kunming 650091, China}

\author{Sheng-Bang Qian}
\affiliation{Department of Astronomy, School of Physics and Astronomy, Yunnan University, Kunming 650091, China}

\author{Si-Yuan Zhu}
\affiliation{School of Physics and Astronomy, Sun Yat-Sen University, Zhuhai, 519082, China}

\author{Fu-Xing Li}
\affiliation{Department of Astronomy, School of Physics and Astronomy, Yunnan University, Kunming 650091, China}

\begin{abstract}
	The origin of the Universe and its material content remains one of the most fundamental questions in science. Gamma-ray bursts (GRBs), with their extreme luminosities and high-redshift detectability, provide a unique window into the history of cosmic formation and chemical evolution. Consequently, the GRB formation rate (FR) has been employed to trace the star formation rate (SFR) across cosmic time. GRBs are conventionally classified into long and short categories (lGRBs and sGRBs) based on their $ T_{90} $ duration. sGRBs are widely employed as tracers of the delayed SFR, owing to their origin linked to the inspiral timescales of compact binary systems. However, some studies suggest that the detection of supernova-associated sGRBs may indicate potential contamination by core-collapse events. In this work, we move beyond the $ T_{90} $ classification and focus exclusively on GRBs with confirmed kilonova signatures, which provide unambiguous evidence of binary compact star mergers, to reassess their connection with the delayed SFR. Through analysis of a kilonova-associated GRB (KN/GRBs) sample, we find that even within this robust subset, the KN/GRB FR displays a trend contrary to that of the delayed SFR at low redshifts ($ z < 1 $). This result challenges the conventional theory by indicating that low-redshift KN/GRBs may not accurately trace the delayed SFR, independent of core-collapse contamination, while further validation with larger KN/GRB samples is essential to determine the reliability of compact binary mergers as probes of delayed SFR.

\end{abstract}

\keywords{Gamma-ray bursts(629); Star formation(1569); Luminosity function(942); }

\section{Introduction} \label{sec:intro}
Gamma-ray bursts (GRBs) are among the most energetic astrophysical events in the universe, characterized by intense bursts of high-energy photons. Based on the bimodal distribution of GRBs in the duration--spectral hardness diagram \citep{1993ApJ...413L.101K}, GRBs are generally classified into two categories: long-duration, soft-spectrum GRBs (lGRBs) and short-duration, hard-spectrum GRBs (sGRBs). These two populations are widely believed to originate from distinct progenitor systems. lGRBs are commonly associated with the core collapse of Wolf--Rayet stars \citep{1993ApJ...405..273W,1998ApJ...494L..45P, 1999ApJ...524..262M}, supported by direct observational evidence linking lGRBs to supernovae (SN; \citealp{2003ApJ...591L..17S, 2004ApJ...609..952Z,2024A&A...684A.164K}). sGRBs are predominantly associated with the mergers of compact binary systems, such as binary neutron stars (BNS) or neutron star–black hole (NS–BH) binaries \citep{1986ApJ...308L..43P, 1989Natur.340..126E, 1992ApJ...395L..83N}. The detection of GW170817 and its electromagnetic counterparts—GRB 170817A and the optical transient AT 2017gfo—provided compelling observational evidence that BNS mergers can give rise to sGRBs \citep{2017ApJ...848L..13A}. Subsequently, continued observations of AT 2017gfo led to its identification as a kilonova (KN) \citep{2017Natur.551...64A,2017ApJ...848L..12A,2017Natur.551...80K}.
KN are transient astronomical events powered by the radioactive decay of heavy r-process elements synthesized in the predominantly isotropic ejecta produced during the merger of BNS or NS–BH systems. The high-energy decay photons are trapped within the expanding ejecta, leading to thermalization and the production of broadband optical–infrared radiation similar to that of a supernova  (SN), but with peak luminosities intermediate between those of classical novae and core-collapse SN\citep{1998ApJ...507L..59L,2005ApJ...634.1202R,2017LRR....20....3M}.


GRBs are often considered tracers of the cosmic star formation rate (SFR), as they originate from the remnants of massive stars. These compact remnants form shortly after the birth of the most massive stars, during their final evolutionary stages, linking GRBs to episodes of active star formation \citep{1998MNRAS.294L..13W,1997ApJ...486L..71T,2000MNRAS.312L..35B,2012A&A...539A.113E}. However, in the case of compact binary mergers, a time delay between the SFR and the merger rate is expected due to the orbital inspiral timescale \citep{2004JCAP...06..007A,2006AIPC..836...58P,2015MNRAS.448.3026W}. As a result, sGRBs are increasingly viewed as potential tracers of delayed SFR.

The intrinsic durations ($T_{90}/(1+z)$) of GRB 080913 \citep{2009ApJ...693.1610G} and GRB 090423 \citep{2009Natur.461.1254T,2009Natur.461.1258S} are less than $2\,s$, yet evidence suggests their origins are consistent with massive stars \citep{2009ApJ...703.1696Z}.
\citet{2013ApJ...764..179B} suggested that a substantial fraction, up to approximately 40\%, of sGRBs detected by \textit{Swift} may originate from collapsars rather than compact binary mergers, indicating contamination of the canonical short GRB population. \citet{2024ApJ...963L..12P} found that a subset of lGRBs with higher event rates exhibit a tendency to align with a delayed SFR model. This correlation leads them to propose that low-redshift lGRBs may, in part, originate from the mergers of compact binary systems. \citet{2025ApJ...987L..13D} proposed event rates of sGRBs can be described by both the delay and undelayed SFR model, suggesting that a subset of sGRBs may originate from the core collapse of massive stars rather than from compact binary mergers. 

Although the correlations between lGRBs - SN and  sGRBs - KN provide strong support for the collapsar and compact binary merger models, respectively, a number of exceptions have emerged that complicate this simple classification. 
Notably, the lGRBs 060614 \citep{2006Natur.444.1044G} and 060505 \citep{2006Natur.444.1047F} lack detectable SN signatures but instead show features consistent with KN emission. 
Likewise, sGRB 200826A ($T_{90}=1.16$\,s) was found to be accompanied by a SN \citep{2021NatAs...5..917A,2021NatAs...5..911Z}, challenging the canonical expectation for sGRB. 
Moreover, recent studies have confirmed KN associations with several lGRBs---GRB 060505 \citep{2021arXiv210907694J}, GRB 080503A \citep{2023ApJ...943..104Z}, GRB 211211A \citep{2022Natur.612..232Y}, and GRB 230307A \citep{2023ApJ...953L...8W}---which are now believed to originate from mergers of compact binary systems.

These findings collectively suggest that a fraction of sGRBs may originate from core-collapse events. Moreover, evidence is emerging that compact binary mergers, traditionally associated with sGRBs, may also contribute to a subset of lGRBs. This indicates that the progenitor channels of both sGRBs and lGRBs are more interconnected and complex than previously thought. As a result, the traditional classical method is no longer universally valid. The link between sGRBs and compact binary mergers should be reconsidered. This undermines the reliability of using $T_{90}$-based sGRBs as tracers of the delay SFR. To address this issue, this paper proposes that GRBs directly associated with kilonovae (KN/GRBs) should be used as more reliable tracers of the cosmic delay SFR. By focusing on this subset of GRBs, we can obtain a cleaner progenitor population, minimizing contamination from core-collapse events. This approach offers a more robust foundation for utilizing GRBs in cosmological studies.

This paper is organized as follow. In Section 2, we introduce sample selection and data analysis method. In Section 3, we show the result of luminosity function (LF) and FR for 19 KN/GRBs. The conclusion and discussion are displayed in Section 4. Throughout the paper, we assume a flat $\Lambda $ universe with ${\Omega _m} =0.3$ and ${H_0} =$ 70 $km\,{s^{ - 1}}Mp{c^{ - 1}}$.

\section{Sample Selection and Data Analysis method} \label{sec:sample}
\subsection{Sample Selection}
We collected a sample that have been confirmed to be associated with KN from the literature, of which 15 were drawn from \citet{2023MNRAS.524.1096L}, while the remaining sources were selected from various previously published studies. This sample includes multi-band observational data, light curve characteristics, and redshift measurements, providing a valuable statistical basis for investigating the properties of KN and their connection to host galaxies.
\citet{2021arXiv210907694J} reported a thermal-like optical radiation component in GRB 060505, which can be interpret as a blue KN.
GRB 211211A was discovered in likely association with a galaxy at z=0.076 \citep{2022hst..prop16923R}. \citet{2023MNRAS.521..269Z} found the light-curve behaviour of GRB 070707 can be explained as an external forward shock afterglow component plus a KN, \citet{2024Natur.626..742Y} reported a lanthanide-rich KN in the aftermath of GRB 230307A. \citet{2025ApJ...979..159S} provided evidence for KN Light for GRB 191019A by joint of an afterglow plus a KN model revealed a better match than an afterglow-only scenario, the resulting KN properties resemble those of AT2017gfo associated with the binary neutron star merger GW170817. Our sample is enlarge to 20 KN/GRBs used to determine the LF and FR. Combining main spectral data from literature and the Gamma-ray Burst Coordinates Network\footnote{\url{https://www.mpe.mpg.de/~jcg/grbgen.html}}. 

Table \ref{tab:1} provided the parameters information, including name, duration $T_{90}$, redshift , low-high energy power-law index  $\alpha$ and $\beta$, peak energy $E_p$ of the $vfv$
spectrum in the observer’s frame, peak flux F, energy band, bolometric luminosity, and references of KN/GRBs. Figure~\ref{fig:1} (a) displays the $T_{90}$ duration distribution for our KN/GRBs sample. Notably, 8 bursts exhibit durations longer than the conventional 2\,s threshold. This observation, combined with the known population of long-duration ($T_{90} > 2\,\text{s}$) events originating from compact binary mergers (as discussed in previous sections), further demonstrates the limitations of using $T_{90}$ as a definitive progenitor classification criterion.

\subsection{Luminosity calculate}
The bolometric peak luminosity $L$ of a GRB are calculate by using:
\begin{equation}
	L = 4 \pi d_L^2(z) F K,
	\tag{3}
\end{equation}
where $d_L(z)$ is the luminosity distance at redshift $z$, defined as:
\begin{equation}
	d_L(z) = \frac{c}{H_0}(1+z) \int_0^z \frac{dz'}{\sqrt{1 - \Omega_m + \Omega_m (1 + z')^3}},
	\tag{4}
\end{equation}
$F$ is the peak flux observed between energy ranges ($E_{\rm min}$, $E_{\rm max}$), and $K$ is the K-correction factor, since the peak fluxes of GRBs are observed over a wide range of redshifts, corresponding to different rest-frame energy bands, we need to make K-correction transform the observed band of telescope in to 1-$10^4$ keV band to obtain the bolometric luminosity of GRBs \citep{2001AJ....121.2879B}. If the $F$ is in units of erg cm$^{-2}$ s$^{-1}$, then the parameter $K$ is defined as:
\begin{equation}
	K = \frac{\int_{1~\mathrm{keV}/(1+z)}^{10^4~\mathrm{keV}/(1+z)} E f(E) \, dE}{\int_{E_{\rm min}}^{E_{\rm max}} E f(E) \, dE}.
	\tag{5}
\end{equation}
If the flux $F$ is in units of photons cm$^{-2}$ s$^{-1}$, then the parameter $K$ is defined as:
\begin{equation}
	K = \frac{\int_{1~\mathrm{keV}/(1+z)}^{10^4~\mathrm{keV}/(1+z)} E f(E) \, dE}{\int_{E_{\rm min}}^{E_{\rm max}} f(E) \, dE}.
	\tag{6}
\end{equation}
$f(E)$ denotes the spectral model of GRBs. there are two spectral models are employed to fit the spectra of GRBs: a power law with an cutoff exponential model (CPL; \citealp{2008ApJS..175..179S}) and the Band model \citep{1993ApJ...413..281B}. The functional forms of these models are as follows:
The CPL model can be expressed as
\begin{equation}
	f(E) = A \left( \frac{E}{100\, \text{keV}} \right)^{\alpha} \exp \left( -\frac{(2 + \alpha) E}{E_p} \right),
	\tag{1}
\end{equation}
where $A$ is the normalization factor, $\alpha$ is the power-law index, and  $E_{\rm p}$ is the peak energy in observer frame. The GRBs are best fitted with Band model can be written as:
\begin{equation}
	f(E) = 
	\begin{cases}
		A \left( \frac{E}{100~\mathrm{keV}} \right)^{\alpha} \exp \left( -\frac{(2 + \alpha) E}{E_{\rm p}} \right), & E < \frac{(\alpha - \beta) E_{\rm p}}{2 + \alpha}, \\
		A \left( \frac{E}{100~\mathrm{keV}} \right)^{\beta} \exp \left[ (\beta - \alpha) \left( \frac{(\alpha - \beta) E_{\rm p}}{(2 + \alpha) 100~\mathrm{keV}} \right)^{\alpha - \beta} \right], & E \geq \frac{(\alpha - \beta) E_{\rm p}}{2 + \alpha},
	\end{cases}
	\tag{2}
\end{equation}
where $A$ is the normalization factor, $\alpha$ and $\beta$ are the low- and high-energy photon indices, respectively. It should be noted that the bursts in our sample were detected by instruments aboard different satellites. To ensure that all bursts in the sample are above the flux limit, we adopt the best sensitivity of $F_{\rm limit} = 1 \times 10^{-8} \, \text{erg cm}^{-2} \text{s}^{-1}$ for KN/GRBs as the flux threshold for the entire sample \citep{2012MNRAS.423.2627W,2025APJL}. Hence, the corresponding luminosity limit at redshift $z$ can be calculated using ${L_{\rm limit}} = 4\pi d_L^2(z) F_{\rm limit}$.

\subsection{Delay time model} \label{sec:delay sfr method}
With the binary merger scenario as a guiding framework, a rate function that accounts for a time delay relative to the SFR was constructed \citep{1999ApJ...511...41T,2004JCAP...06..007A,2006AIPC..836...58P,2015MNRAS.448.3026W}. 
This delayed rate is formulated as the convolution between a given SFR model and a delay-time distribution, $f(\Delta t)$, which represents the probability distribution of time delays relative to the SFR. The intrinsic KN/GRBs rate is expressed as the convolution of the SFR with the delay-time distribution $f(\Delta t)$.
\begin{equation}
	R_{\mathrm{sGRB}}(z) \propto \int_{z_{min}}^{\infty} \mathrm{SFR}(z') \left(f(t(z) - t(z')) \frac{\mathrm{d}{\tau}}{\mathrm{d}z'}\right) \, \mathrm{d}z' \, .
\end{equation}
where $ z_{\min}(z)$  is obtained on solving $ t_z - t_{z_{\min}} = \tau_{\min} = 10~\text{Myr}$ \citep{2018MNRAS.477.4275P}.

Here, we use the SFR model of \citet{2014ARA&A..52..415M} , which can be expressed as
\begin{equation}
	SFR(z) = \frac{(1+z)^{2.7}}{1 + \left(\frac{1+z}{2.9}\right)^{5.6}},
\end{equation}
We consider three models for the time delay \citep{2015ApJ...812...33S}:

(1) Gaussian merger delay time-scale model.The probability intensity distribution as
\begin{equation}
	f(\tau) = \frac{1}{\sigma\sqrt{2\pi}} \exp\left(-\frac{(\tau - \tau_0)^2}{2\sigma^2}\right),
\end{equation}
where $\tau_0$= 2 Gyr, and $\sigma$ = 0.3 Gyr.

(2) Log-normal merger delay time-scale model
\begin{equation}
	f(\tau) \, d\ln\tau = \frac{1}{\sigma\sqrt{2\pi}}\exp\left(-\frac{(\ln\tau - \ln \tau _{0})^2}{2\sigma^2}\right)   \, d\ln\tau,
\end{equation}
where $\tau_0$= 2.9 Gyr, and $\sigma$ = 0.2 Gyr.

(3) Power-law merger delay time-scale model
\begin{equation}
	f(\tau) = \tau^{-\alpha_{\tau}},
\end{equation}
where $\alpha_{\tau}$=0.81.  $ \tau = t(z)-t(z')$ is the delay time between the formation and the merger of binary star systems. The $ t(z)$ is age of the Universe
calculated in a flat $\Lambda $ universe \citep{2024ApJ...969..123L}, which is calculate as 
\begin{equation}
	t(z) = H_0^{-1} \int_{z}^{\infty} \left[ (1 + z') \sqrt{\Omega_m (1 + z')^3 + \Omega_\Lambda} \right]^{-1} \, dz',
\end{equation}
we plot the distribution of KN/GRBs rate based on the above three model in Figure \ref{fig:5}.

\subsection{Lynden-Bell's ${c^ - }$ method} \label{sec:c- method}


Several selection effects influence the observed redshift distribution of GRBs \citep{2007NewAR..51..539C}, and consequently their inferred event rate. 
The most significant of these arises from the observational limitations of satellite. These satellite has a flux limit, meaning that it is unable to detect GRBs fainter than a given threshold. As a result, the observed sample is truncated, and the intrinsic redshift distribution of GRBs cannot be reliably determined without first accounting for these observational biases.

\citet{1971MNRAS.155...95L} (Lynden-Bell's ${c^ - }$) putted forward a non-parametric methods that can be used to study the LF and density evolution for quasar sample. But \citet{1992scma.conf..173P} pointed that the drawback of non-parametric method is \textit{ad hoc} assumption of \textit{uncorrelated variables}. To overcoming this defects, \citet{1992ApJ...399..345E} (EP) developed a new method (EP-L method) to test whether luminosity (L) and redshift (z) are correlated or not. This method is used on a variety of cosmological objects, such as quasar \citep{1971MNRAS.155...95L,1992ApJ...399..345E}, Active galactic nucleus \citep{1999ApJ...518...32M,2014ApJ...786..109S,2021ApJ...913..120Z}, GRBs \citep{2015ApJS..218...13Y,2015ApJ...806...44P,2025ApJ...978..160L}, and FRB \citep{2019JHEAp..23....1D,2024ApJ...973L..54C}. In this paper, we also use this method to drive the cosmological evolution of LF and FR of our sample.

\begin{figure*}[t]
	\centering
	\includegraphics[width=0.45\columnwidth]{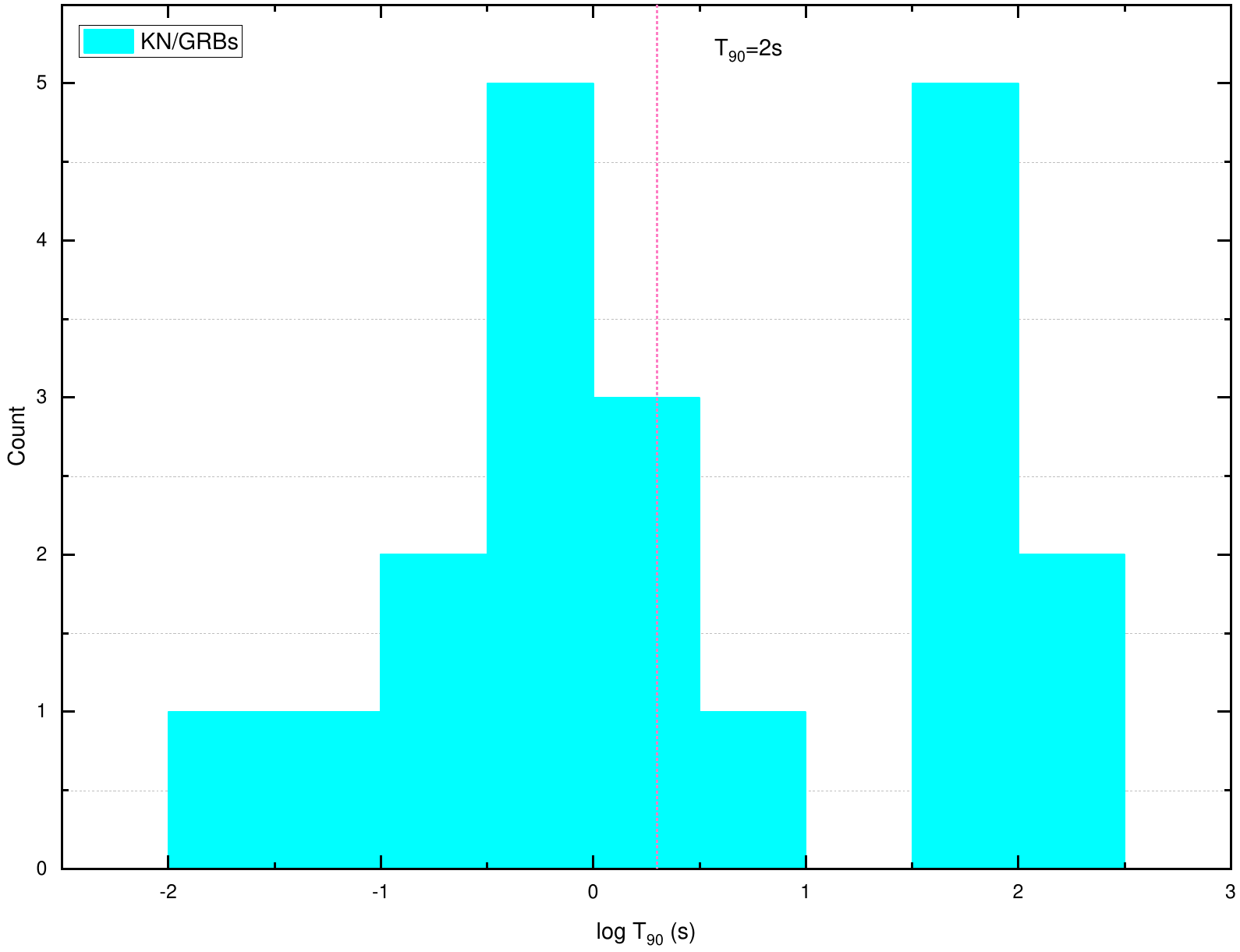}
	\includegraphics[width=0.45\columnwidth]{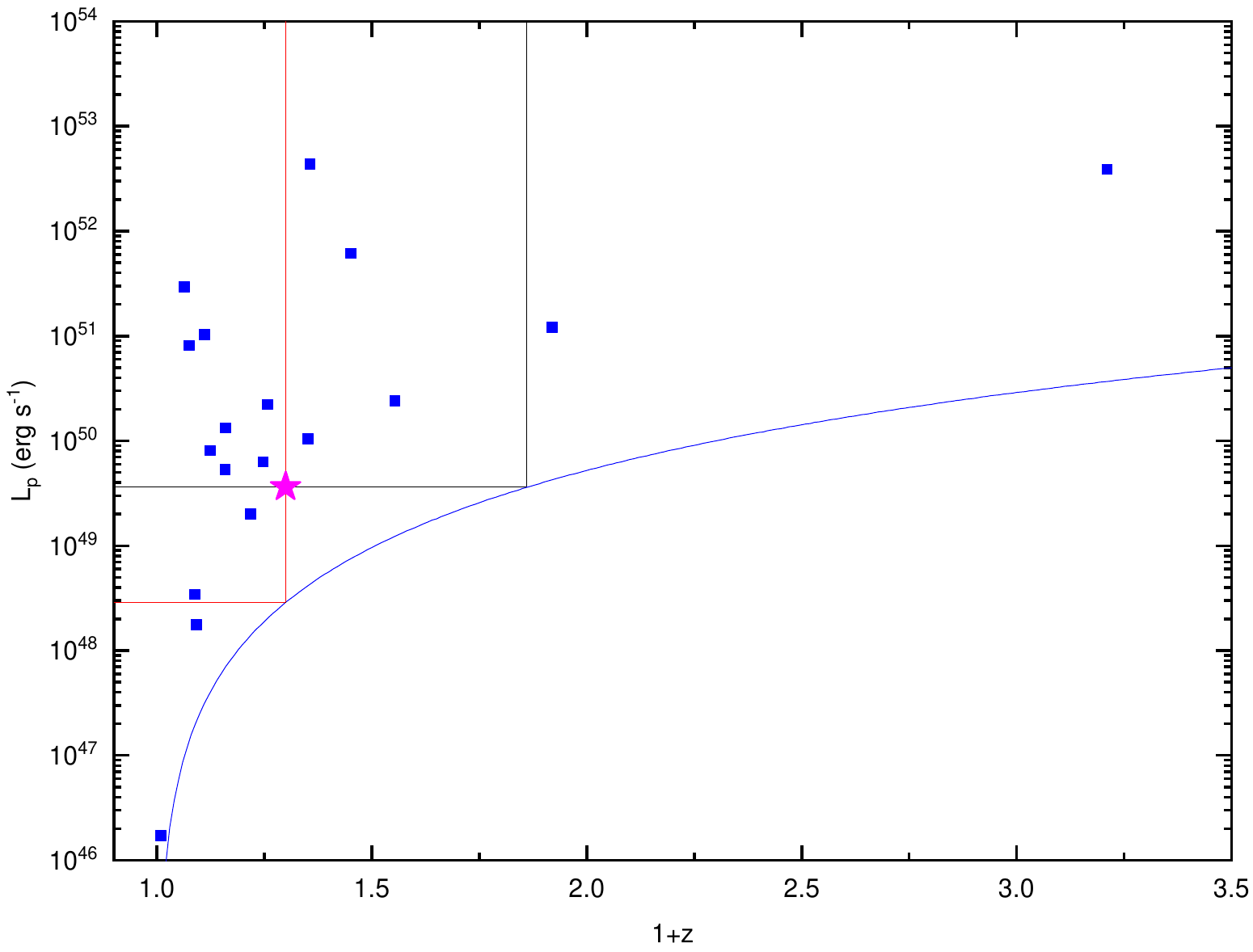}
	\includegraphics[width=0.45\columnwidth]{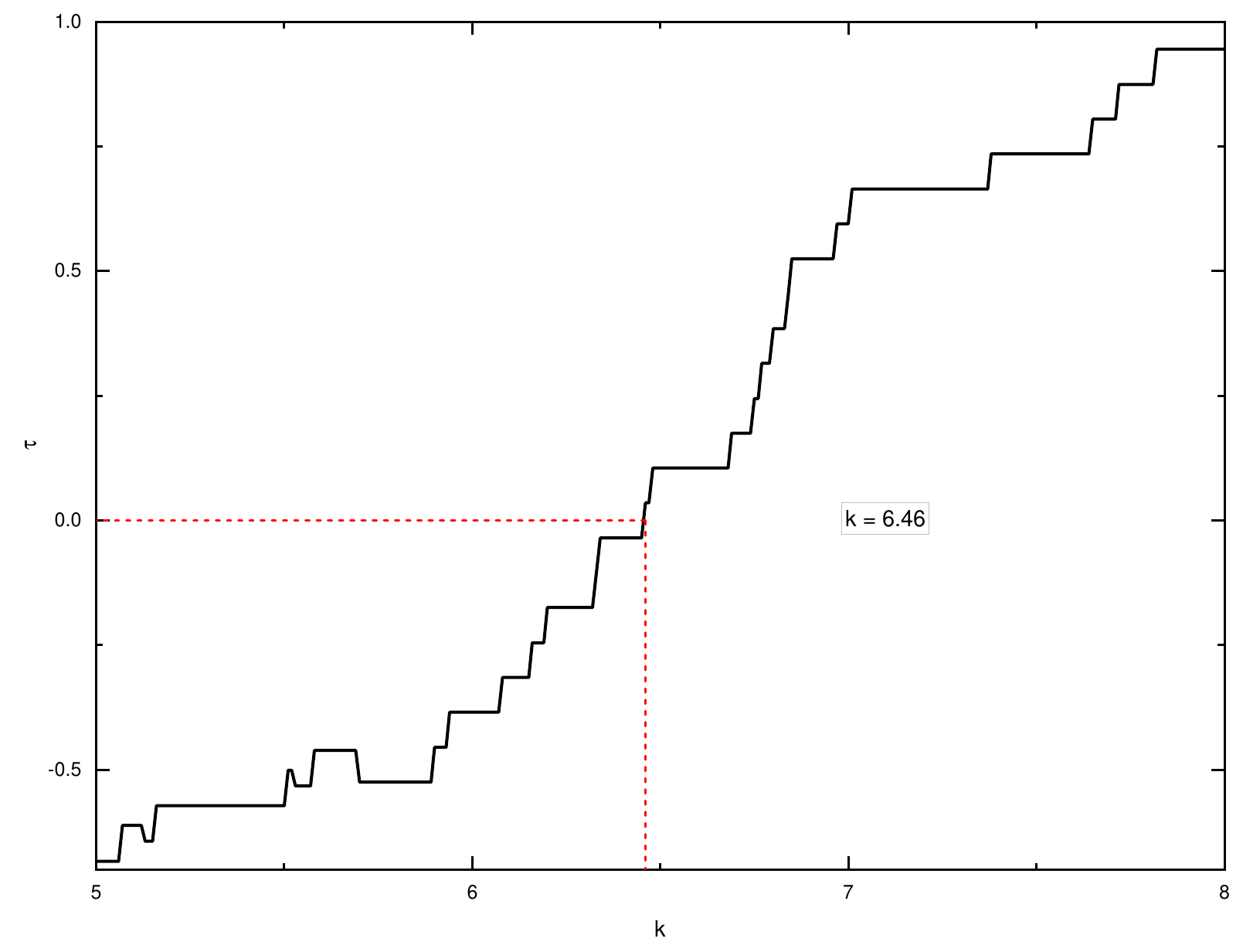}
	\includegraphics[width=0.45\columnwidth]{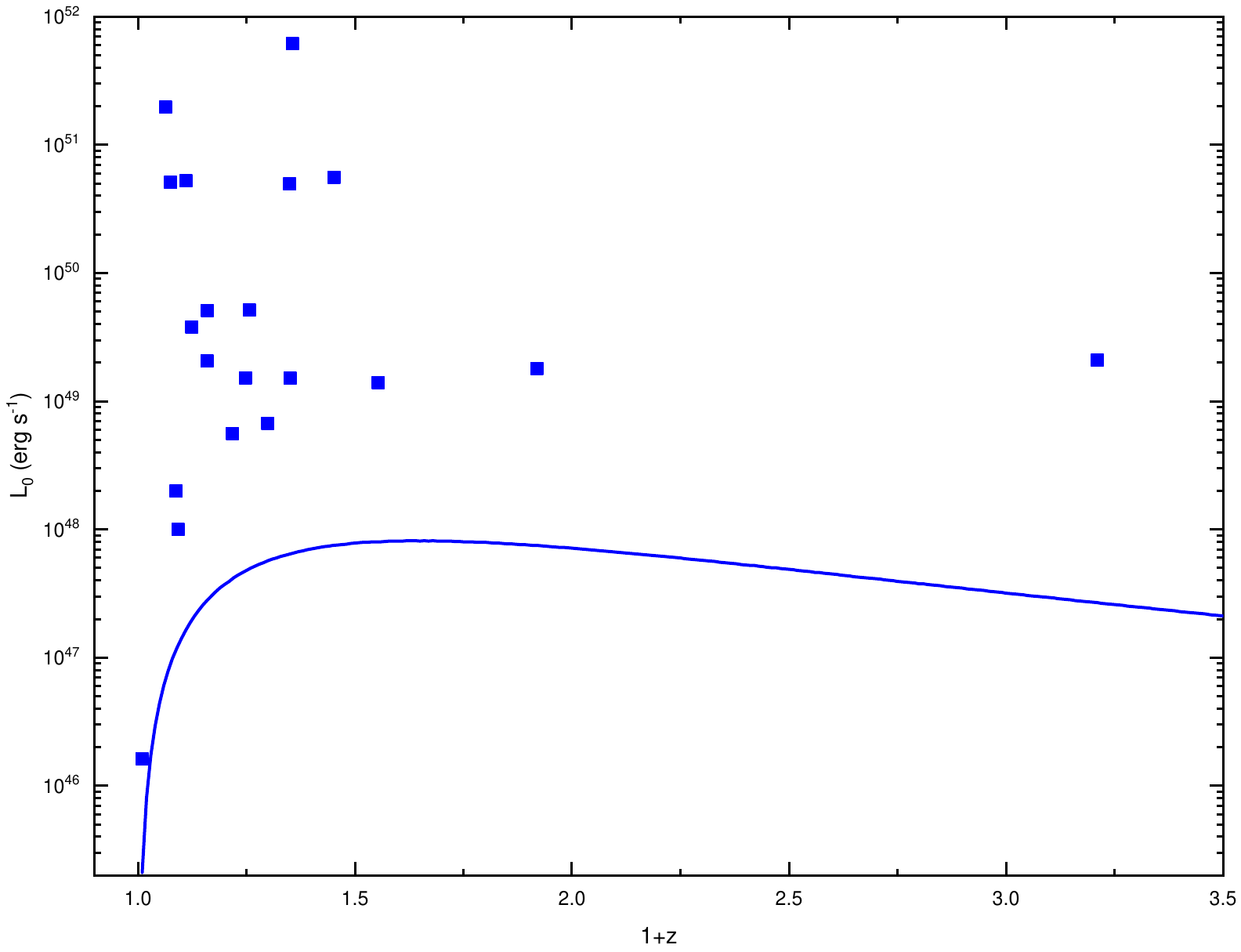}
	\caption{Distributions and correlations for the KN/GRBs sample: 
		Left top: Duration ($T_{90}$) distribution of KN/GRBs events; 
		Right top: Bolometric luminosity distribution where individual points represent different KN/GRBs, with the line indicating the sensitivity limit of $1.0 \times 10^{-8}\,\mathrm{erg\,cm^{-2}\,s^{-1}}$; 
		Left bottom: In the Kendall $\tau$ correlation test, the red dotted line represents the null hypothesis ($\tau = 0$), and the measured correlation strength of $k = 6.46$ suggests that the evolutionary dependence between luminosity and redshift has been effectively removed; 
		Right bottom: De-evolved luminosity function following $L = L_{0}(1 + z)^{6.46}$ for our sample of 20 KN/GRBs, removing the redshift evolution component.}
	\label{fig:1}
\end{figure*}

In Figure \ref{fig:1} (b), it is easy see that luminosity and redshift are dependent. Therefore, we need to eliminate this effect at first by using EP-L method \citep{1992ApJ...399..345E}. Luminosity evolves with function ${g_k}(z) = {(1 + z)^k}$, we can determine the value of $k$. As pointed as previous author \citep{1992ApJ...399..345E,2015ApJS..218...13Y,2019JHEAp..23....1D}, we use the kendall $\tau $ statistical method to drive the $k$.
Figure \ref{fig:1} (b) show the distribution of luminosity and redshift, for a random point $i$ (${z_i},L{}_i $) in this plane, the ${J_{\rm{i}}}$ can be defined as
\begin{equation}
	{J_{\rm{i}}} = \{ {\rm{j}}|{L_{\rm{j}}} \ge {L_i},{z_j} \le z_i^{\max }\}
\end{equation}
where $L_i$ is the luminosity of $i$th KN/GRB and $z_i^{\max }$ is the maximum redshift at which the KN/GRBs with the luminosity limit. The range of ${J_{\rm{i}}}$ is shown as the black rectangle in Figure \ref{fig:1} (b). The number of KN/GRBs that contained in this region is defined as ${n_i}$. The number ${N_i}={n_i-1}$, which means takeing the $i$th point out. Similarly, the $J_i^{'}$ can be defined as
\begin{equation}
	J_i^{'} = \{ {\rm{j}}|{L_{\rm{j}}} \ge L_{_i}^{\lim },{z_j} \le {z_i}\}
\end{equation}
where $L_{_i}^{\lim }$ is the limiting corrected gamma-ray luminosity at redshift ${z_i}$. In Figure \ref{fig:1} (b), the range of $J_i^{'}$ is shown as red rectangle. The number of KN/GRBs that contained in this range is defined as ${M_i}$.

Firstly, we consider to define the number of KN/GRBs that have redshift $z$ less than or equal to ${z_i}$ in black rectangle as ${R_i}$. The statistic $\tau$ is \citep{1992ApJ...399..345E}
\begin{equation}
	\tau  = \frac{{\sum\nolimits_i {({R_{\rm{i}}} - {E_i})} }}{{\sqrt {\sum\nolimits_i {{V_{\rm{i}}}} } }}
\end{equation}
where ${E_i} = \frac{{1 + ni}}{2}$ and ${V_i} = \frac{{n_i^2 - 1}}{{12}}$ are the expected mean and the variance for the hypothesis of independence, respectively. 
If $R_i$ is exactly uniformly distributed between 1 and $n_i$, then the samples with $R_i \leq E_i$ and $R_i \geq E_i$ should be approximately equal in number, and the test statistic $\tau$ will be close to zero. If we choose a functional form of $g(z) = {(1 + z)^k}$ that make the test statistic $\tau = 0$, then the effect of luminosity evolution can be removed by applying the transformation $L_0 = L / g(z)$. Figure \ref{fig:1} (d) show the distribution of non-evolving gamma-ray luminosity. The $k$ value is 6.46 for 20 KN/GRBs, as shown in Figure \ref{fig:1} (c) .

After eliminate the effect of the luminosity evolution, we can obtain the cumulative luminosity function with non-parametric method from following equation
\begin{equation}
	\psi ({L_{0{\rm{i}}}}{\rm{) = }}\mathop \prod \limits_{{\rm{j}} < i} (1 + \frac{1}{{{N_j}}})    
	\label{equ:5}
\end{equation}
and the cumulative number distribution can be derived from
\begin{equation}
	\phi (z{\rm{) = }}\mathop \prod \limits_{j < i} (1 + \frac{1}{{{M_j}}})           
	\label{equ:6}
\end{equation}

The differential form of $\phi (z)$ represents the formation rate of $\rho $, which can be expressed as
\begin{equation}
	\rho (z) = \frac{{d\phi (z)}}{{dz}}(1 + z){\left( {\frac{{dV(z)}}{{dz}}} \right)^{ - 1}}           
	\label{equ:7}
\end{equation}
where ${\frac{{dV(z)}}{{dz}}}$ represents the differential comoving volume, which can be written as
\begin{eqnarray}
	\begin{split}
		\frac{{dV(z)}}{{dz}} =4\pi \left( {\left. {\frac{c}{{{H_0}}}} \right)} \right.\frac{{D_L^2}}{{{{(1 + z)}^2}}} \times \frac{1}{{\sqrt {1 - {\Omega _m} + {\Omega _m}{{(1 + z)}^3}} }} 
		\label{equ:8}
	\end{split}
\end{eqnarray}

\section{Result} 
\label{sec:luminosity function and formation rate of KN/GRB}
\begin{figure}[ht!]  
	\centering
	\includegraphics[width=3.5in]{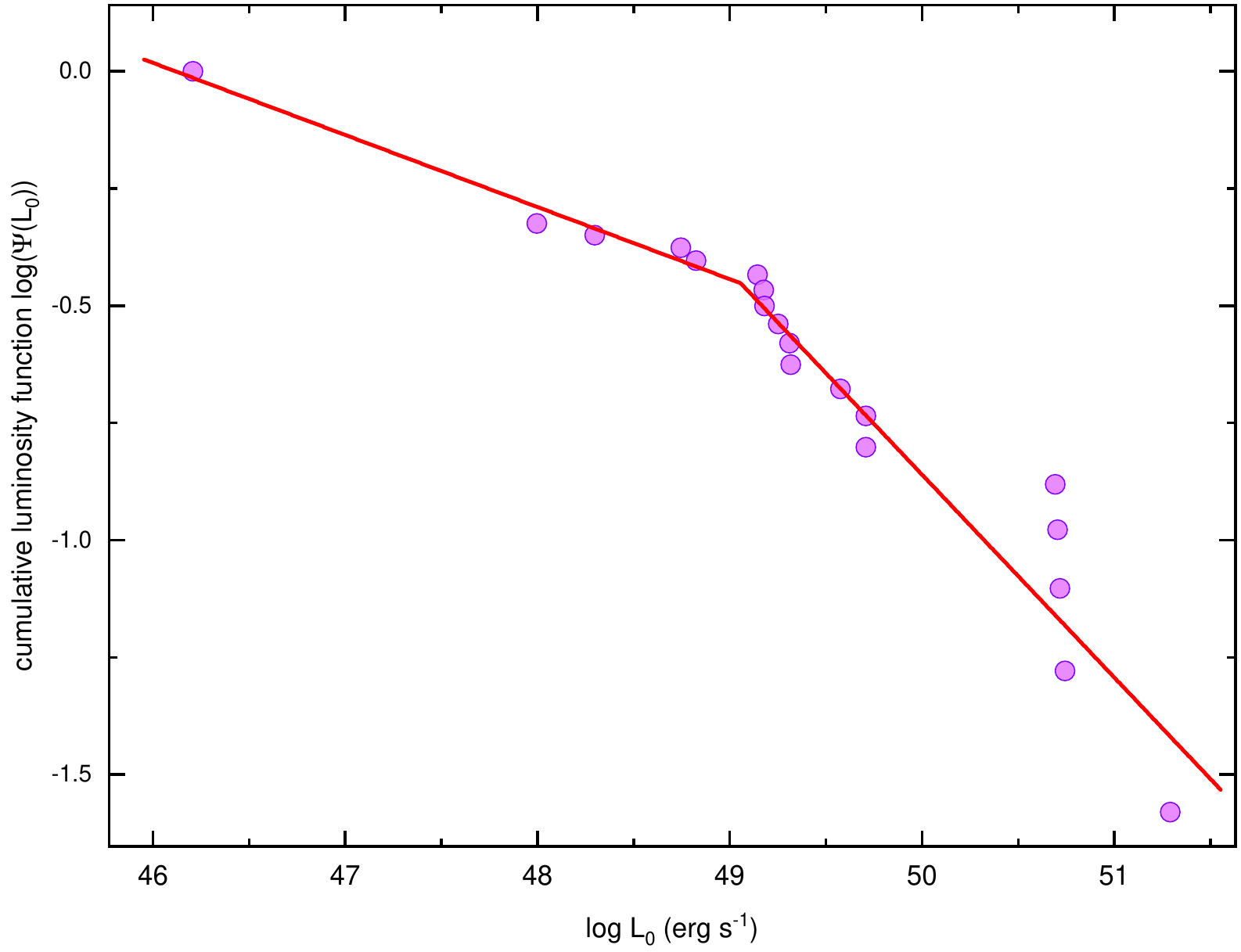}
	\includegraphics[width=3.5in]{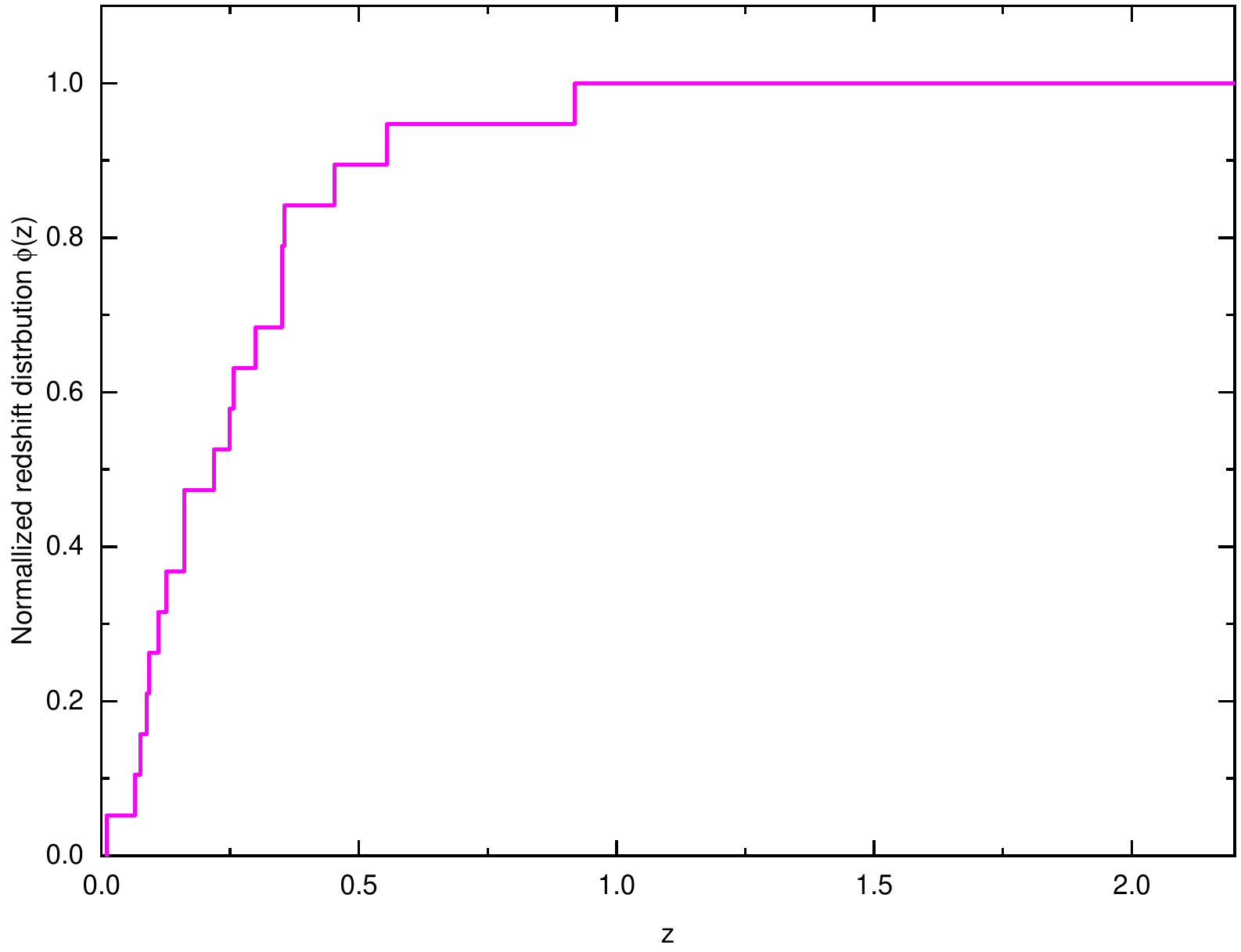}
	\caption{Left: The distribution of cumulative luminosity function $\psi({L_0})$, which is normalized to unity at the lowest luminosity. The function is fitted by red solid line with broken power law. The luminosity function can be expressed as $\psi ({L_0}) = L_0^{-0.15 \pm 0.04},{L_0} < L_0^b$ and $\psi ({L_0}) = L_0^{-0.43 \pm 0.03},{L_0} > L_0^b$; Right: Normalized cumulative redshift distribution.}
	\label{fig:2}
\end{figure}

In this section, we present the derived LF and FR for our KN/GRBs sample. Following the methodology described in previous sections, we quantify the luminosity evolution through a non-parametric approach using the Kendall $\tau$ correlation test.
We use broken power law to fit the  cumulative luminosity function in Figure \ref{fig:2} (a), and the following results were obtained:
\begin{eqnarray}
	\psi ({L_0}) \propto \left\{ {\begin{array}{*{20}{c}}
			{L_0^{ -0.15 \pm 0.04},{L_0} < L_0^b}\\
			{L_0^{-0.43 \pm 0.03},{L_0} > L_0^b}                    
	\end{array}} \right.
	\label{equ:9}
\end{eqnarray}\\
where $L_0^b = (1.15 \pm 0.85) \times 10^{49} \, \text{erg}\, \text{s}^{-1}$ is the break point. It is worth to point out that the $\psi (L/g(z))$ is local luminosity at $z=0$ for the luminosity evolution is removed, and the luminosity function can be rewrite as $\psi ({L_0}) = \psi {(L/(1 + z)^{6.46})}$. Therefore, the break luminosity at z can be deduced $L_z^b = L_0^b{(1 + z)^{6.46}}$.

\begin{figure}[ht!]  
	\includegraphics[width=5in]{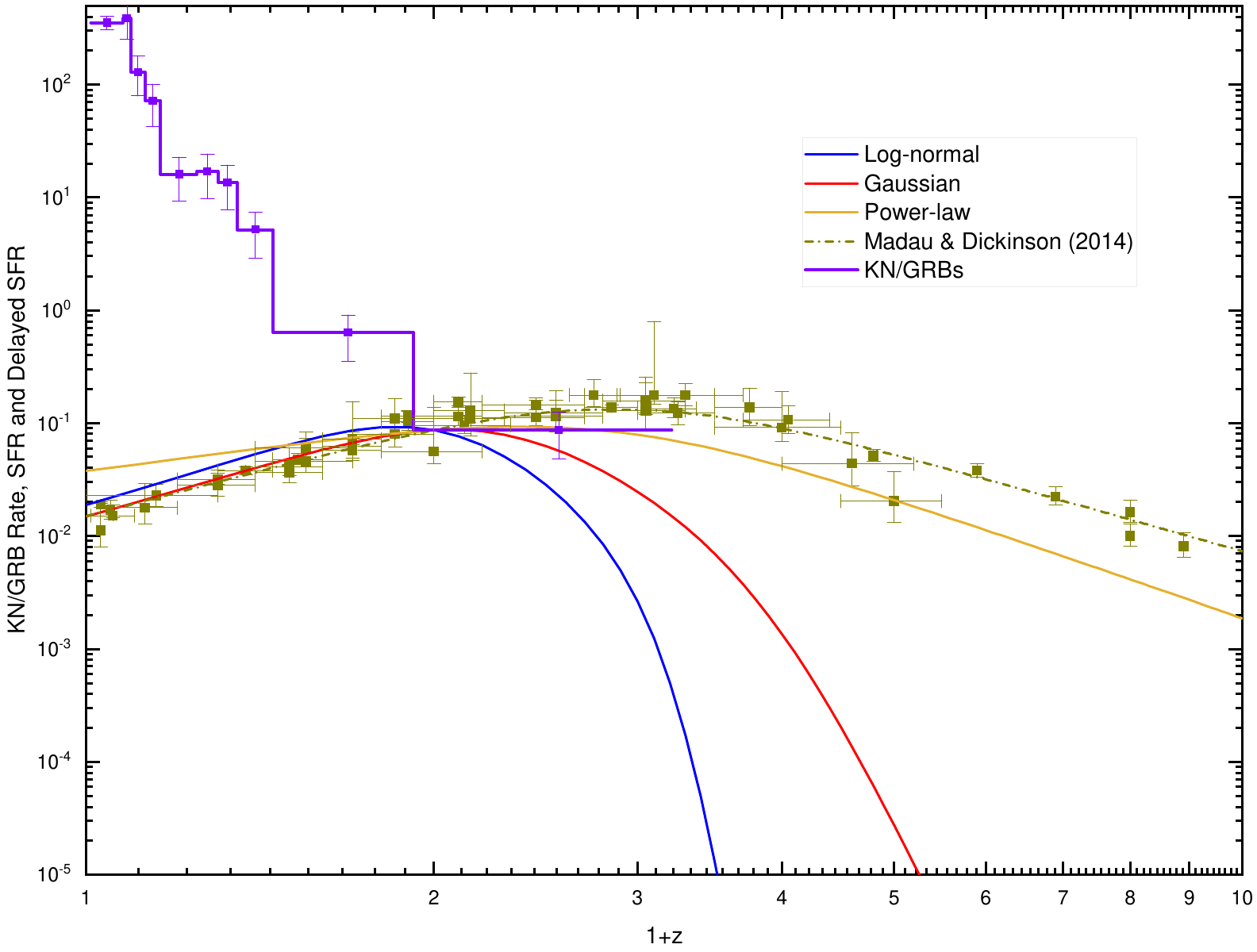}
	\centering
	\caption{Comparison KN/GRBs rate and SFR. The thick colored curves display the best fits achieved by applying merged delay models to each distribution. The SFR data and fit line were collected from \citet {2014ARA&A..52..415M} (Brown dots). The blue line are the formation rate of fitting 20 KN/GRBs sources. The error bar gives a 1 $\sigma$ poisson error \citep{1986ApJ...303..336G}. These fitting lines are normalized at z=1.}
	\label{fig:5}
\end{figure}

Figure~\ref{fig:2} (b) displays the cumulative redshift distribution $\phi(z)$ for our sample. Prior to deriving the FR, we first obtain the differential form $d\phi(z)/dz$ through numerical differentiation. The error bar gives a 1 $\sigma$ poisson error \citep{1986ApJ...303..336G}. As evident in Figure~\ref{fig:5}, the resulting FR $\rho(z)$ exhibits a pronounced decreasing trend with increasing redshift. When compared to the SFR and delay SFR, our results shows low redshifts excess ($z< 1$).


\section{DISCUSSIONS AND CONCLUSIONS }

In this work, we compile a sample of 20 GRBs that have been securely associated with KN from the literature. one would expect KN/GRBs — being directly connected to the binary compact object merger — to follow the delay SFR more closely than the general sGRB population. Utilizing Lynden-Bell's $C^{-}$ method to eliminate the redshift-luminosity correlation, we derive the intrinsic LF and FR of 20 KN/GRBs. This enables us to perform a robust comparison between the KN/GRBs event rate and the delay SFR. Our key findings are as follows:
(1) A Kendall $\tau$ test yields a statistically significant correlation with a $k$-value of 6.46 for the 20 KN/GRBs in our sample;
(2) The luminosity function distributions of KN/GRBs can be fitted by a smoothly broken power-law (BPL) function. The broken luminosities are around $1.15 \times 10^{49} \, \text{erg} \, \text{s}^{-1}$.
(3) We observe a decrease trend in the KN/GRBs event rate at low redshifts ($z < 1$).
These results suggest that the observed KN/GRBs event rate can't trace SFR and delay SFR. and may instead point to intrinsic differences in the GRB progenitor population or its dependence on environmental factors.


It is well established that GRBs originate from two primary channels: the mergers of compact binaries \citep{1986ApJ...308L..43P} or the core collapse of massive stars \citep{1993ApJ...405..273W}. The conventional paradigm holds that a subset of lGRBs arises from core-collapse events, often accompanied by supernova-like emission signatures \citep{1998Natur.395..670G, 2003Natur.423..847H}. Due to their association with massive star formation, lGRBs are widely regarded as powerful probes of the cosmic SFR at high redshifts.
However, previous studies have revealed a discrepancy between the observed lGRB FR and the SFR, with some suggesting contamination from compact binary merger events \citep{2015ApJS..218...13Y, 2015ApJ...806...44P, 2024ApJ...963L..12P}. Intriguingly, \citet{2025APJL} calculated the event rate of GRBs associated with supernovae and still found a significant excess at low redshifts. This suggests that the low-redshift excess cannot be fully explained by progenitor channel mixing alone. Our findings also offer additional support for this viewpoint.
Further evidence for complexity in the GRB population comes from luminosity-dependent studies. \citet{2023ApJ...958...37D} found that the event rates of low-luminosity lGRBs exceed the SFR, whereas high-luminosity lGRBs are in excellent agreement with the cosmic star formation history. Similarly, \citet{2025MNRAS.542..215L} noted that the rates of medium and faint GRBs exceed the SFR at low redshifts, while the rate of bright GRBs closely follows the SFR.
These findings indicate that the low-redshift excess in the GRB rate is likely driven by luminosity-dependent effects or environmental factors, rather than being solely attributable to the distinction between compact binary mergers and core-collapse supernovae. 


Extra explanations for this discrepancy include the small size of the current KN/GRBs sample. The current sample is deficient in high-redshift ($z>1$) KN/GRBs, with only one detected events in the $z>2$ range, leading to considerable uncertainty. Future observations will be crucial for exploring the relationship between the KN/GRB FR and the delayed SFR in the high-redshift universe.

\section*{Acknowledgements}
This work was supported by the Postdoctoral Fellowship Program of CPSF under Grant Number GZC20252095, the China Postdoctoral Science Foundation under Grant Number 2025M773194, Caiyun Postdoctoral Program in Yunnan Province of China (grant No. C615300504124), the National Natural Science Foundation of China (grant Nos. 12503040, 11933008 and 12303040), National Key R\&D Program of China (grant No. 2022YFE0116800), and Yunnan Fundamental Research Projects (grant NOs. 202501AS070055, 202503AP140013, 202201AT070092 and 202401AT070143).

\bibliographystyle{apj}
\bibliography{ms}

\clearpage

\clearpage

\setlength{\tabcolsep}{1mm}{
	\renewcommand\arraystretch{1}
	\begin{center}
		\begin{longtable}{lccccccccc}
			\caption{20 KN/GRBs Included in Our analysis}
			\label{tab:1} \\
			\hline%
			GRB &${T_{90}}$  & z & $\alpha$  &$\beta$  &  $\textit{E}_{\rm p}$   & ${E_{\min}}$-${E_{\max}}$ &F  & $L_p$ &ref \\
			&   (s)         &    &  &  &(keV)      & (keV)   &  (ph/cm$^{2}$/s)   & $erg\, s^{-1}$ &    \\
			(1)&(2)& (3) &(4) & (5) &(6) & (7)&(8)&(9)&(10)\\
			\hline%
			\endhead%
			\hline%
			\endfoot%
			\hline%
			\endlastfoot%
			050709&0.07&0.1606&-0.7&-&$ 83 ^{+ 18 }_{- 18 }$&30-400&$12.1 ^{+ 0.4 }_{- 0.4 }$&$ 1.33  ^{+ 0.04  }_{- 0.04  }\times 10 ^{ 50 }$&\citet{2018PASP..130e4202Z}\\
			050724&96&0.257&-2.02&-&$ 78.91 ^{+ 8 }_{- 8 }$&15-150&$3.35 ^{+ 0.31 }_{- 0.31 }$&$ 2.23  ^{+ 0.21  }_{- 0.21  }\times 10 ^{ 50 }$&\citet{2018PASP..130e4202Z}\\
			060505&4&0.089&-0.627&-&$ 124.925 $&15-150&$1.29 ^{+ 0.2 }_{- 0.24 }$&$ 3.43  ^{+ 0.53  }_{- 0.64  }\times 10 ^{ 48 }$&swift website\\
			060614A&108&0.125&-1.82&-&$ 134.1 ^{+ 24.2 }_{- 24.2 }$&15-150&$11.4 ^{+ 0.7 }_{- 0.7 }$&$ 8.05  ^{+ 0.49  }_{- 0.49  }\times 10 ^{ 49 }$&\citet{2018PASP..130e4202Z}\\
			061201&0.76&0.111&-0.3&-&$ 873 ^{+ 458 }_{- 284 }$&20-3000&$3.19 ^{+ 0.72 }_{- 2.72 }\times 10 ^{ -5a }$&$ 1.03  ^{+ 0.23  }_{- 0.88  }\times 10 ^{ 51 }$ &\citet{2006GCN..5890....1G}\\
			070707&1&0.35&-0.57&-&$ 427 ^{+ 374 }_{- 144 }$&20-2000&$8.1 ^{+ 2.9 }_{- 6.7 }\times 10 ^{ -6a }$&$ 3.41  ^{+ 1.22  }_{- 2.82  }\times 10 ^{ 51 }$&\citet{2007GCN..6615....1G}\\
			070714B&64&0.92&-0.88&-&$ 81.62 ^{+ 36.76 }_{- 36.76 }$&15-150&$2.75 ^{+ 0.16 }_{- 0.16 }$&$ 1.21  ^{+ 0.07  }_{- 0.07  }\times 10 ^{ 51 }$&\citet{2025ApJ...987L..13D}\\
			070809&1.39&0.2187&-1.33&-&$ 145.5 ^{+ 70 }_{- 70 }$&15-150&$0.97 ^{+ 0.12 }_{- 0.12 }$&$ 2.00  ^{+ 0.25  }_{- 0.25  }\times 10 ^{ 49 }$&\citet{2018PASP..130e4202Z}\\
			080503&170&0.3&-1.23&-&$ 226.61 $&15-150&$0.67 ^{+ 0.13 }_{- 0.13 }$&$ 3.63  ^{+ 0.71  }_{- 0.71  }\times 10 ^{ 49 }$&\citet{2025ApJ...987L..13D}\\
			100625A&0.33&0.452&-0.1&-&$ 414 ^{+ 128 }_{- 78 }$&20-2000&$8.1 ^{+ 1.5 }_{- 1.5 }\times 10 ^{ -6a }$&$ 6.16  ^{+ 1.14  }_{- 1.14  }\times 10 ^{ 51 }$&\citet{2010GCN.10890....1G}\\
			111117A&0.47&2.211&-0.69&-&$ 370 ^{+ 37 }_{- 37 }$&15-150&$2.8 ^{+ 0.2 }_{- 0.2 }$&$ 3.89  ^{+ 0.28  }_{- 0.28  }\times 10 ^{ 52 }$&\citet{2018PASP..130e4202Z}\\
			130603B&0.18&0.3564&-0.73&-&$ 660 ^{+ 100 }_{- 100 }$&20-10000&$1 ^{+ 0.2 }_{- 0.2 }\times 10 ^{ -4a }$&$ 4.37  ^{+ 0.87  }_{- 0.87  }\times 10 ^{ 52 }$&\citet{2013GCN.14771....1G}\\
			140903A&0.3&0.351&-1.36&-&$ 44.17 ^{+ 16.2 }_{- 16.2 }$&15-150&$2.5 ^{+ 0.2 }_{- 0.2 }$&$ 1.05  ^{+ 0.08  }_{- 0.08  }\times 10 ^{ 50 }$&\citet{2018PASP..130e4202Z}\\
			150101B&0.018&0.093&-0.58&-&$ 96.55 $&15-150&$0.73 ^{+ 0.3 }_{- 0.3 }$&$ 1.76  ^{+ 0.72  }_{- 0.72  }\times 10 ^{ 48 }$&\citet{2025ApJ...987L..13D}\\
			160821B&0.48&0.16&-1.37&-&$ 84 ^{+ 19 }_{- 19 }$&10-1000&$9.16 ^{+ 1.19 }_{- 1.19 }$&$ 5.32  ^{+ 0.69  }_{- 0.69  }\times 10 ^{ 49 }$&\citet{2016GCN.19843....1S}\\
			170817A&2&0.009783&-0.89&-&$ 82 ^{+ 21 }_{- 21 }$&8-1000&$1.9 ^{+ 0.2 }_{- 0.2 }$&$ 1.72  ^{+ 0.18  }_{- 0.18  }\times 10 ^{ 46 }$&GCN Circular 21520\\
			191019A&64.35&0.248&-1.99&-&$ 1.19 ^{+ 0.2 }_{- 0.2 }$&15-150&$2.15 ^{+ 0.06 }_{- 0.06 }$&$ 6.33  ^{+ 0.18  }_{- 0.18  }\times 10 ^{ 49 }$&swift website\\
			200522A&0.62&0.554&-0.53&-&$ 77.76 $&15-150&$2.02$&$ 2.40 \times 10 ^{ 50 }$&swift website\\
			211211A&34.3&0.076&-1.3&-2.4&$ 646.8  ^{+ 7.8 }_{- 7.8 }$&10-1000&$324.9  ^{+ 1.5 }_{- 1.5 }$&$ 8.12  ^{+ 0.04  }_{- 0.04  }\times 10 ^{ 50 }$&\citet{2021GCN.31210....1M}\\
			230307A&35 &0.0646&-1.07&-&$ 936 ^{+ 3 }_{- 3 }$&10-1000&$791 ^{+ 4 }_{- 4 }$&$ 2.93  ^{+ 0.01  }_{- 0.01  }\times 10^{ 51 }$&\citet{2023GCN.33411....1D}\\

		\end{longtable}
		\begin{tablenotes}
			\textbf{Note}: References for the spectral parameters ($\textit{E}_{\rm p}$, $\alpha $, $\beta $, peak flux , ${E_{\max}}$, and redshift of GRB 080503, GRB 070707: \url{ https://www.mpe.mpg.de/~jcg/grbgen.html} ; $a$: These peak fluxes are in units of erg cm$^{-2}$  s$^{-1}$. b: The redshift of GRB 070707 and GRB 080503 were assumed as 0.35 and 0.3 \citep{2008A&A...486..405M,2023ApJ...943..104Z}. The data of GRB 060505, 191019A and GRB 200522A were acquired from The Swift/BAT Gamma-Ray Burst website: \url{https://swift.gsfc.nasa.gov/results/batgrbcat/}.
		\end{tablenotes}
	\end{center}

\end{document}